\documentclass
[nofootinbib,superscriptaddress,notitlepage,a4paper,onecolumn]{revtex4-1}%
\usepackage{amssymb}
\usepackage{amsfonts}
\usepackage{amsmath}
\usepackage{graphicx}
\usepackage{color}
\usepackage{version}
\usepackage{numinsec}
\usepackage{hyperref}%
\providecommand{\U}[1]{\protect \rule{.1in}{.1in}}

\begin{document}
\title{Phase-space approach to polaron response: Kadanoff and
Feynman-Hellwarth-Iddings-Platzmann re-examined}
\author{Dries Sels}
\email{Corresponding author: dries.sels@uantwerpen.be}
\author{Fons Brosens}
\email{fons.brosens@uantwerpen.be}
\affiliation{Physics Department, University of Antwerp, Universiteitsplein 1, 2060
Antwerpen, Belgium}

\begin{abstract}
A method is presented to obtain the linear response coefficients of a system
coupled to a bath. The method is based on a systematic truncation of the
Liouville equation for the reduced distribution function. The first order
truncation results are expected to be accurate in the low temperature and weak
coupling regime. Explicit expressions for the conductivity of the Fr\"{o}hlich
polaron are obtained, and the discrepancy between the Kadanoff and the
Feynman-Hellwarth-Iddings-Platzmann mobility is elucidated.

\end{abstract}
\maketitle

\section{Introduction}

Since its inception, the mobility of the Fr\"{o}hlich
polaron~\cite{cit:Frohlich1937,cit:Frohlich1950,cit:Frohlich1954} has been the
subject of many theoretical studies. For an excellent in-depth overview and
discussion we refer to a textbook by Alexandrov and
Devreese~\cite{cit:AlexandrovDevreese2010} and to lecture notes by
Devreese~\cite{cit:DevreeseVarenna}. A prominent approach was proposed by
Feynman \textit{et al.}~\cite{cit:FHIP1962} (hereafter referred to as FHIP),
based on the path-integral formalism. This method is nonperturbative in the
sense that no expansion in the coupling constant is assumed, but it is limited
to first order in the applied electric field. However, in the asymptotic limit
of weak electron-phonon coupling and low temperature, the FHIP polaron
mobility differs by a factor of $3/\left(  2\hbar \beta \omega_{LO}\right)  $
--with $\omega_{LO}$ the dispersionless longitudinal optical phonon frequency,
and $\beta=1/\left(  k_{B}T\right)  $ where $k_{B}$ is Boltzmann's constant
and $T$ is the temperature-- from the mobility which
Kadanoff~\cite{cit:Kadanoff1963} found later on from the Boltzmann equation
within the relaxation time approximation. As already pointed out in FHIP, the
same factor of $3/\left(  2\hbar \beta \omega_{LO}\right)  $ appears in
comparison with earlier
results~\cite{cit:HowarthSondheimer1953,cit:LowPines1955,cit:Osaka1961}. It
has been argued in~\cite{cit:PeetersDevreese1983,cit:PeetersDevreese1984} that
this discrepancy might be due to interchanging two limits (with both the
frequency of the applied electric field and the electron-phonon coupling
strength tending to zero). But this mathematical argument implicitly assumes
the Kadanoff result to be valid, which we dispute.

In the present paper, we propose an alternative approach, based on the
dynamics of the Wigner distribution function~\cite{cit:Wigner1932}. The
methodology is basically inspired by the Feynman-Vernon influence
functionals~\cite{cit:FeynmanVernon1963}, rather than on Feynman's variational
path integral treatment of the ground state energy of the
polaron~\cite{cit:Feynman1955}. However, instead of considering the path
integral for the wave function of a system, we contributed
in~\cite{cit:SBMinfluencefunctional2010} to a path integral description of the
Wigner distribution function. Concentrating on a particle that linearly
interacts with a set of independent harmonic oscillators, the influence
functional for the Wigner distribution function could be reduced to a double
path integral in the path variables of the particle, if the oscillators are
initially in thermodynamical equilibrium. In a subsequent
paper~\cite{cit:SBreducedLiouville2013} we derived a perturbation series for
the propagator of the reduced Wigner function (i.e., the Wigner function for
the particle of interest). By exactly resumming this series, we found a Dyson
integral equation for the reduced propagator, from which the equation of
motion for the reduced Wigner function could be derived. For general
temperature and interaction strength, the resulting equation with a dressed
propagator is still under investigation. We here concentrate on linear
response at weak coupling and low temperature, in order to elucidate the
discrepancy between the FHIP and the Kadanoff mobility.

The paper is organized as follows. In section~\ref{sec:Generic} we extract the
assumptions and results from the
papers~\cite{cit:SBMinfluencefunctional2010,cit:SBreducedLiouville2013} which
are relevant for our present purpose. In section \ref{sec:Linear} we present
an approximate, however systematically improvable, truncation method to derive
the linear response coefficients from the equation of motion for the reduced
Wigner function. We present a detailed discussion on the conductivity of the
Fr\"{o}lich polaron in section \ref{sec:Frohlich}, after which we conclude in
\ref{sec:Conclusion}. Supplementary information on the used truncation scheme
is provided in appendix \ref{sec:AppTruncated}. Additional calculations on the
relaxation time approximation and on FHIP are found in appendix
\ref{sec:AppendixRelaxation} and \ref{sec:AppendixFHIP} respectively.

\section{Reduced Wigner function for a generic polaron
system\label{sec:Generic}}

Consider the following generic polaron Hamiltonian
\begin{equation}
H=\frac{\mathbf{p}^{2}}{2m}-e\mathbf{E}\left(  t\right)  \cdot \mathbf{x}%
+\sum_{k}\hbar \omega_{k}\left(  b_{\mathbf{k}}^{\dagger}b_{\mathbf{k}}%
+\frac{1}{2}\right)  +\sum_{\mathbf{k}}\left(  \gamma \left(  k\right)
\exp \left(  -i\mathbf{k\cdot x}\right)  b_{\mathbf{k}}^{\dagger}+\gamma^{\ast
}\left(  k\right)  \exp \left(  i\mathbf{k\cdot x}\right)  b_{\mathbf{k}%
}\right)  , \label{eq:Generic:Hamiltonian}%
\end{equation}
where $\left(  \mathbf{x,p}\right)  $ represent the electron coordinate and
momentum operator. It is coupled to some bosonic field $b_{\mathbf{k}}$ in a
isotropic translational invariant way, i.e. $\gamma(k)=\gamma(\mathbf{k}%
)=\gamma(\left \vert \mathbf{k}\right \vert )$. Also the phonon frequency
$\omega_{k}=\omega_{\left \vert \mathbf{k}\right \vert }$ is isotropic. The
electron is subject to a time dependent but homogeneous electric field
$\mathbf{E(}t\mathbf{).}$

Because the system is translational invariant in the absence of the field
$\mathbf{E(}t\mathbf{),}$ we suppose that the electron distribution is
homogeneous, and that the phonon bath is initially in thermal equilibrium:%
\begin{equation}
f\left(  \mathbf{r,p,}\left \{  x_{\mathbf{k}},p_{\mathbf{k}}\right \}
,t=-\infty \right)  =f\left(  \mathbf{p},t=-\infty \right)  \prod_{\mathbf{k}%
}\frac{\tanh \frac{\beta \hbar \omega_{k}}{2}}{\pi \hbar}\exp \left(  -\frac
{\tanh \frac{\beta \hbar \omega_{k}}{2}}{\hbar \omega_{k}}\left(  \frac
{p_{\mathbf{k}}^{2}}{m}+m\omega_{k}^{2}x_{\mathbf{k}}^{2}\right)  \right)  .
\label{eq:Generic:f_initial}%
\end{equation}
Knowledge of the (reduced) Wigner distribution function $f\left(
\mathbf{p},t\right)  $ would allow to calculate the current density, and hence
the conductivity $\sigma$
\begin{align}
\mathbf{J}(t)  &  =\frac{e}{m}\int \mathbf{p}f\left(  \mathbf{p},t\right)
\mathrm{d}\mathbf{p}\label{eq:Generic:defineJ}\\
&  =\int_{-\infty}^{t}\sigma(t-t^{\prime})\mathbf{E(}t^{\prime}\mathbf{)}%
\mathrm{d}t^{\prime}. \label{eq:Generic:definesigma}%
\end{align}
In general, $\sigma$ is a tensor but, due to the cylindrical symmetry of
(\ref{eq:Generic:Hamiltonian}), it becomes diagonal. The Wigner-Liouville
equation for the case of a phonon bath which initially is in thermal
equilibrium, and for a general potential $V\left(  \mathbf{x,}t\right)  ,$ was
derived in a recent paper~\cite{cit:SBreducedLiouville2013}. For the
electronic Hamiltonian $\frac{\mathbf{p}^{2}}{2m}-e\mathbf{E}\left(  t\right)
\cdot \mathbf{x}$ under consideration here, the relevant equations~(I.2--I.4)
of Ref.~\cite{cit:SBreducedLiouville2013} simplify into%
\begin{multline}
\left(  \frac{\partial}{\partial t}+e\mathbf{E}\left(  t\right)  \cdot \frac
{d}{d\mathbf{p}}\right)  f\left(  \mathbf{p},t\right)  =\sum_{\mathbf{k}}%
\frac{2\left \vert \gamma \left(  k\right)  \right \vert ^{2}}{\hbar^{2}%
}\label{eq:Generic:Liouville}\\
\times%
{\displaystyle \iiint}
\Theta \left(  t^{\prime}\leq t\right)  \left(
\begin{array}
[c]{c}%
\left(
\begin{array}
[c]{c}%
\left(  n_{B}\left(  \omega_{k}\right)  +1\right)  \cos \left(  \mathbf{k\cdot
}\left(  \mathbf{x}-\mathbf{x}^{\prime}\right)  +\omega_{k}\left(
t-t^{\prime}\right)  \right) \\
+n_{B}\left(  \omega_{k}\right)  \cos \left(  \mathbf{k\cdot}\left(
\mathbf{x}-\mathbf{x}^{\prime}\right)  -\omega_{k}\left(  t-t^{\prime}\right)
\right)
\end{array}
\right) \\
\times \left(  K_{0}\left(  \mathbf{x},\mathbf{p-}\frac{\hbar \mathbf{k}}%
{2},t|\mathbf{x}^{\prime},\mathbf{p}^{\prime}\mathbf{+}\frac{\hbar \mathbf{k}%
}{2},t^{\prime}\right)  -K_{0}\left(  \mathbf{x},\mathbf{p+}\frac
{\hbar \mathbf{k}}{2},t|\mathbf{x}^{\prime},\mathbf{p}^{\prime}\mathbf{+}%
\frac{\hbar \mathbf{k}}{2},t^{\prime}\right)  \right)
\end{array}
\right)  f\left(  \mathbf{p}^{\prime},t^{\prime}\right)  \mathrm{d}t^{\prime
}\mathrm{d}\mathbf{x}^{\prime}\mathrm{d}\mathbf{p}^{\prime},
\end{multline}%
\begin{align}
\text{with }K_{0}\left(  \mathbf{x},\mathbf{p},t|\mathbf{x}^{\prime
},\mathbf{p}^{\prime},t^{\prime}\right)   &  =\delta \left(  \mathbf{p}%
-\mathbf{p}^{\prime}-\int_{t^{\prime}}^{t}e\mathbf{E}\left(  \sigma \right)
\, \mathrm{d}\sigma \right)  \delta \left(  \mathbf{x}-\mathbf{x}^{\prime}%
-\frac{\mathbf{p}^{\prime}}{m}\left(  t-t^{\prime}\right)  -\int_{t^{\prime}%
}^{t}\frac{e\mathbf{E}\left(  \sigma \right)  }{m}\left(  t-\sigma \right)
\, \mathrm{d}\sigma \right)  ,\label{eq:Generic:K0}\\
n_{B}\left(  \omega_{k}\right)   &  =1/\left(  e^{\beta \hbar \omega_{k}%
}-1\right)  . \label{eq:Generic:nB}%
\end{align}
Note that we have dropped the position dependence of the distribution function
because both the initial state (\ref{eq:Generic:f_initial}) and the electric
field are homogeneous. In the absence of the electric field, the time
evolution of the Wigner distribution becomes%
\begin{equation}
\frac{\partial f_{\mathbf{E=}0}\left(  \mathbf{p},t\right)  }{\partial t}%
=\sum_{\mathbf{k}}\frac{2\left \vert \gamma \left(  k\right)  \right \vert ^{2}%
}{\hbar^{2}}\int_{0}^{\infty}\left(
\begin{array}
[c]{c}%
\left(
\begin{array}
[c]{c}%
\left(  n_{B}\left(  \omega_{k}\right)  +1\right)  \cos \left(  \left(
\frac{\left(  \mathbf{p}+\hbar \mathbf{k}\right)  ^{2}-\mathbf{p}^{2}}{2m\hbar
}-\omega_{k}\right)  s\right) \\
+n_{B}\left(  \omega_{k}\right)  \cos \left(  \left(  \frac{\left(
\mathbf{p}+\hbar \mathbf{k}\right)  ^{2}-\mathbf{p}^{2}}{2m\hbar}+\omega
_{k}\right)  s\right)
\end{array}
\right)  f_{\mathbf{E=}0}\left(  \mathbf{p+}\hbar \mathbf{k},t-s\right) \\
-\left(
\begin{array}
[c]{c}%
\left(  n_{B}\left(  \omega_{k}\right)  +1\right)  \cos \left(  \left(
\frac{\left(  \mathbf{p}+\hbar \mathbf{k}\right)  ^{2}-\mathbf{p}^{2}}{2m\hbar
}+\omega_{k}\right)  s\right) \\
+n_{B}\left(  \omega_{k}\right)  \cos \left(  \left(  \frac{\left(
\mathbf{p}+\hbar \mathbf{k}\right)  ^{2}-\mathbf{p}^{2}}{2m\hbar}-\omega
_{k}\right)  s\right)
\end{array}
\right)  f_{\mathbf{E=}0}\left(  \mathbf{p},t-s\right)
\end{array}
\right)  \mathrm{d}s. \label{eq:Generic:Liouville_E=0}%
\end{equation}
It seems unlikely that this integro-differential equation can be solved in
closed form. Even a stationary solution $f_{\mathbf{E=}0}^{\text{stat}}\left(
\mathbf{p}\right)  $ in the absence of an electric field obeys a non-trivial
integral equation. Using $\int_{0}^{\infty}\cos \left(  as\right)
\mathrm{d}s=\pi \delta \left(  a\right)  $, some elementary algebra reveals
that, within the continuum limit, it satisfies the balance equation%
\begin{equation}
\int \Pi \left(  \mathbf{p}+\hbar \mathbf{k}\rightarrow \mathbf{p}\right)
f_{\mathbf{E=}0}^{\text{stat}}\left(  \hbar \mathbf{k+p}\right)  d\mathbf{k}%
=f_{\mathbf{E=}0}^{\text{stat}}\left(  \mathbf{p}\right)  \Pi \left(
\mathbf{p}\right)  , \label{eq:Generic:f_E=0_stat}%
\end{equation}
where we adopt an analogous notation as introduced by Devreese and
Evrard~\cite{cit:DevreeseEvrard1976}, and define%
\begin{align}
\Pi \left(  \mathbf{p}+\hbar \mathbf{k}\rightarrow \mathbf{p}\right)   &
=\frac{V\left \vert \gamma \left(  k\right)  \right \vert ^{2}}{\left(
2\pi \right)  ^{2}\hbar}\left(
\begin{array}
[c]{c}%
\left(  n_{B}\left(  \omega_{k}\right)  +1\right)  \delta \left(  \frac{\left(
\mathbf{p}+\hbar \mathbf{k}\right)  ^{2}-\mathbf{p}^{2}}{2m}-\hbar \omega
_{k}\right) \\
+n_{B}\left(  \omega_{k}\right)  \delta \left(  \frac{\left(  \mathbf{p}%
+\hbar \mathbf{k}\right)  ^{2}-\mathbf{p}^{2}}{2m}+\hbar \omega_{k}\right)
\end{array}
\right)  ,\label{eq:Generic:PI(...)_define}\\
\Pi \left(  \mathbf{p}\right)   &  =\int \Pi \left(  \mathbf{p}\rightarrow
\mathbf{p}+\hbar \mathbf{k}\right)  d\mathbf{k}.
\label{eq:Generic:PI_total_define}%
\end{align}
Even this equation is hard to solve in its generality. One can however check
by straightforward algebra that $f_{\mathbf{E=}0}^{\text{stat}}\left(
\mathbf{p}\right)  \propto \exp \left(  -\beta \mathbf{p}^{2}/2m\right)  $
satisfies Eq.(\ref{eq:Generic:f_E=0_stat}). In order to elucidate the
discrepancy between the mobility results of FHIP and Kadanoff, we limit the
further discussion to linear response at weak coupling and low temperature.

\section{Linear response at weak coupling and low
temperature\label{sec:Linear}}

Limiting the discussion to first order in the electric field and to first
order in $\left \vert \gamma \left(  k\right)  \right \vert ^{2},$ the dependence
on $\mathbf{E}$ of the reduced Wigner propagator~(\ref{eq:Generic:K0}) can be
neglected, and the Wigner-Liouville equation~(\ref{eq:Generic:Liouville})
simplifies into%
\begin{multline}
\left(  \frac{\partial}{\partial t}+e\mathbf{E}\left(  t\right)  \cdot \frac
{d}{d\mathbf{p}}\right)  f\left(  \mathbf{p},t\right)  =\sum_{\mathbf{k}}%
\frac{2\left \vert \gamma \left(  k\right)  \right \vert ^{2}}{\hbar^{2}%
}\label{eq:Linear:Liouville}\\
\times \int_{-\infty}^{t}\left(
\begin{array}
[c]{c}%
f\left(  \mathbf{p+\hbar \mathbf{k}},s\right)  \left(
\begin{array}
[c]{c}%
\left(  n_{B}\left(  \omega_{k}\right)  +1\right)  \cos \left(  \left(
t-s\right)  \left(  \mathbf{k\cdot}\frac{\mathbf{p+}\frac{\hbar \mathbf{k}}{2}%
}{m}-\omega_{k}\right)  \right) \\
+n_{B}\left(  \omega_{k}\right)  \cos \left(  \left(  t-s\right)  \left(
\mathbf{k\cdot}\frac{\mathbf{p+}\frac{\hbar \mathbf{k}}{2}}{m}+\omega
_{k}\right)  \right)
\end{array}
\right) \\
-f\left(  \mathbf{p},s\right)  \left(
\begin{array}
[c]{c}%
\left(  n_{B}\left(  \omega_{k}\right)  +1\right)  \cos \left(  \left(
t-s\right)  \left(  \mathbf{k\cdot}\frac{\mathbf{p+}\frac{\hbar \mathbf{k}}{2}%
}{m}+\omega_{k}\right)  \right) \\
+n_{B}\left(  \omega_{k}\right)  \cos \left(  \left(  t-s\right)  \left(
\mathbf{k\cdot}\frac{\mathbf{p+}\frac{\hbar \mathbf{k}}{2}}{m}-\omega
_{k}\right)  \right)
\end{array}
\right)
\end{array}
\right)  \mathrm{d}s.
\end{multline}
It seems impossible to solve this highly non-Markovian initial value problem exactly.

Here we propose an approach which is inspired by the truncated Wigner
approximation as, e.g., extensively discussed by
Polkovnikov~\cite{cit:Polkovnikov2010}. Its application to general coupling
strength and arbitrary temperature is under current investigation. However,
for sufficiently small electron-phonon coupling strength $\gamma \left(
k\right)  $ and sufficiently low temperature, the truncation after the first
moment is justified, as argued in detail in Appendix~\ref{sec:AppTruncated}.
It results in the following equation of motion~(\ref{eq:AppTruncated:dJ/dt})
for the current density:
\begin{equation}
\frac{d\mathbf{J}(t)}{dt}+\int_{-\infty}^{t}\mathbf{J}(s)\chi(t-s)\mathrm{d}%
s=\frac{e^{2}}{m}\mathbf{E(}t\mathbf{),} \label{eq:Linear:current}%
\end{equation}
where the memory function $\chi$ of the system is given by%
\begin{equation}
\chi(t)=t\sum_{\mathbf{k}}\frac{2\left \vert \gamma \left(  k\right)
\right \vert ^{2}}{3\hbar}\frac{\mathbf{k}^{2}}{m}\left(
\begin{array}
[c]{c}%
\left(  n_{B}\left(  \omega_{k}\right)  +1\right)  \sin \left(  t\left(
\frac{\hbar \mathbf{k}^{2}}{2m}+\omega_{k}\right)  \right) \\
+n_{B}\left(  \omega_{k}\right)  \sin \left(  t\left(  \frac{\hbar
\mathbf{k}^{2}}{2m}-\omega_{k}\right)  \right)
\end{array}
\right)  . \label{eq:Linear:chi}%
\end{equation}
The definition~(\ref{eq:Generic:definesigma}) of the conductivity thus yields
the following relation between the Laplace transform $%
\mathcal{L}%
\left(  \sigma,\Omega \right)  $ of the conductivity and the Laplace transform
$%
\mathcal{L}%
\left(  \chi,\Omega \right)  $ of the memory function:
\begin{equation}%
\mathcal{L}%
\left(  \sigma,\Omega \right)  =\frac{e^{2}}{m}\frac{1}{\Omega+%
\mathcal{L}%
\left(  \chi,\Omega \right)  }, \label{eq:Linear:Laplace_sigma}%
\end{equation}
from which one can, for example, immediately extract the (long wavelength)
optical absorption coefficient~\cite{cit:DevreeseSitterGoovaerts1972}%
\begin{equation}
\Gamma(\omega)=\frac{Z_{0}}{n}\operatorname{Re}\left[
\mathcal{L}%
(\sigma,i\omega)\right]  , \label{eq:Linear:OA_DSG}%
\end{equation}
where $n$ is the crystals refractive index and $Z_{0}=(\epsilon_{0}c)^{-1}$ is
the impedance of free space. Further results of course depend on the specifics
of the system at hand. Here we apply the proposed model to the Fr\"{o}hlich polaron.

\section{Fr\"{o}hlich polaron\label{sec:Frohlich}}

For the optical Fr\"{o}hlich polaron one considers $\omega_{k}=\omega_{LO}$ to
be constant. The coupling%
\begin{equation}
\left \vert \gamma \left(  k\right)  \right \vert ^{2}=\frac{\hbar^{2}\omega
_{LO}^{2}}{\mathbf{k}^{2}}\frac{4\pi \alpha}{V}\sqrt{\frac{\hbar}{2m\omega
_{LO}}} \label{eq:Frohlich:gamma(k)}%
\end{equation}
scales with the dimensionless coupling constant $\alpha.$ Then, in the
continuum limit, the remaining integral in Eq.~(\ref{eq:Linear:chi}) is
Gaussian and results in%
\begin{align}
\chi(t)  &  =\frac{2\alpha \omega_{LO}^{2}}{3\sqrt{2\pi}}\left[  \left(
2n_{B}(\omega_{LO})+1\right)  \frac{\cos \left(  \omega_{LO}t\right)  }%
{\sqrt{\omega_{LO}t}}-\frac{\sin \left(  \omega_{LO}t\right)  }{\sqrt
{\omega_{LO}t}}\right] \label{eq:Frohlich:chi(t)}\\
&  =\frac{\alpha \omega_{LO}^{2}}{3}\left[  \left(  2n_{B}(\omega
_{LO})+1\right)  J_{-1/2}\left(  \omega_{LO}t\right)  -J_{1/2}\left(
\omega_{LO}t\right)  \right]  ,\nonumber
\end{align}
where $J_{\pm1/2}$ denotes the Bessel function of the first kind of order
$\pm1/2.$ The Laplace transform~\cite{cit:Watson1966} of $\chi$ is given by
\begin{equation}%
\mathcal{L}%
\left(  \chi,\Omega \right)  =\frac{\alpha \omega_{LO}}{3\sqrt{\left(
\frac{\Omega}{\omega_{LO}}\right)  ^{2}+1}}\left(  \left(  2n_{B}(\omega
_{LO})+1\right)  \sqrt{\sqrt{\left(  \frac{\Omega}{\omega_{LO}}\right)
^{2}+1}+\frac{\Omega}{\omega_{LO}}}-\sqrt{\sqrt{\left(  \frac{\Omega}%
{\omega_{LO}}\right)  ^{2}+1}-\frac{\Omega}{\omega_{LO}}}\right)  .
\label{eq:Frohlich:chi_Laplace}%
\end{equation}
Consequently the low temperature
DC-conductivity~(\ref{eq:Linear:Laplace_sigma}) is%
\begin{equation}
\sigma_{DC}=\lim_{\Omega \rightarrow0}%
\mathcal{L}%
\left(  \sigma,\Omega \right)  =\frac{3e^{2}}{2\alpha m\omega_{LO}n_{B}%
(\omega_{LO})}\approx \frac{3e^{2}}{2\alpha m\omega_{LO}}e^{\beta \hbar
\omega_{LO}}. \label{eq:Frohlich:sigmaDC}%
\end{equation}
It should immediately be noted that this result differs by a factor of $3$
from that of Kadanoff~\cite{cit:Kadanoff1963} and by a factor of $\left(
2\hbar \beta \omega_{LO}\right)  $ from that of FHIP~\cite{cit:FHIP1962}, i.e.,%
\begin{equation}
\sigma_{DC}=\underset{\text{Kadanoff}}{3~\sigma_{DC}}=2\hbar \beta \omega
_{LO}\underset{\text{FHIP}}{\sigma_{DC}}.
\label{eq:Frohlich:sigma_onze_Kad_FHIP}%
\end{equation}

The result is however in agreement with a prediction made by
Los'~\cite{cit:LOS}, based on a Green's superoperator calculation of Kubo's
formula. It was already argued by FHIP, that in the $\Omega \rightarrow0$ limit
the full Boltzmann equation should be solved in order to get an accurate
result for the DC mobility, an approximate solution of which was later
provided by Kadanoff~\cite{cit:Kadanoff1963}. It was furthermore argued, in
Ref.~\cite{cit:PeetersDevreese1983, cit:PeetersDevreese1984}, that the
$3/\left(  2\hbar \beta \omega_{LO}\right)  $ discrepancy was caused by an
interchange of the $\Omega \rightarrow0$ and $\alpha \rightarrow0$ limit. One
might wonder whether interchanging these limits gives different results in the
current approach. The equation of motion for the current density was obtained
by expanding the scattering term around $p\rightarrow0.$ One might guess that
interchanging the limits by first taking the limit of $\Omega \rightarrow0,$
hence $t\rightarrow \infty,$ and then the limit of $p\rightarrow0$ will result
in a similar difference. As argued in detail in
appendix~\ref{sec:AppendixRelaxation} this is not the case.

The discussion in appendix~\ref{sec:AppendixRelaxation} furthermore
immediately explains the factor of $3$ discrepancy between the present model
and the result of Kadanoff. The in-scattering term in the Boltzmann equation,
expressed in terms of the angular correlation factor
in~\cite{cit:Kadanoff1963}, is completely neglected by Kadanoff and dismissed
as vanishingly small. But neglecting this in-scattering violates particle
number conservation. Within the present approach the in-scattering component
is non-vanishing. The component linear in $E$ exactly subtracts $2/3\left(
2\alpha \omega_{LO}\right)  $ from the inverse scattering rate resulting in a
mobility which is three times higher than the one calculated within the
relaxation time approximation. It is clear that the present approach does not
violate particle number conservation, neither do FHIP and Los'.

The additional $2\hbar \beta \omega_{LO}$ difference with FHIP however remains
to be explained. In appendix~\ref{sec:AppendixFHIP} we reexamine the FHIP
approximation in the language of the distribution function rather than path
integrals for the reduced density matrix. This illuminates the main problem in
the FHIP approximation. First and foremost, unlike what is argued by FHIP, it
is detrimental to assume an initial product state between the bath and the
system for the evolution of the model. Although the true system will quickly
thermalize to the temperature of the bath, the model system of FHIP does not
thermalize, because it is completely harmonic and consequently fully
integrable. In order to obtain a physical trial distribution one must assume
that the complete model system was in thermal equilibrium instead of in a
product state of the system with a thermal bath. Apart from this small change
the analysis in appendix~\ref{sec:AppendixFHIP} is completely in line with
FHIP. The final low temperature DC conductivity however reads%
\[
\sigma_{DC}=\frac{3e^{2}}{2\alpha m^{\ast}\omega_{LO}}e^{\beta \hbar \omega
_{LO}}.
\]
where the effective mass $m^{\ast}/m=v^{2}/w^{2}$ is defined in terms of
Feynman's variational parameters. Since $w\approx v$ and thus $m^{\ast}\approx
m$ for sufficiently small $\alpha,$ we recover the same
result~(\ref{eq:Frohlich:sigmaDC}) as derived by our linearized equation of
motion. It is clear that the present FHIP reanalysis does not have the
spurious $2\hbar \beta \omega_{LO}$ terms.

\section{Conclusion\label{sec:Conclusion}}

In conclusion we have presented a method to obtain the conductivity of a
generic polaron. In the low temperature and weak coupling regime a truncation
after the first moment is justified and the conductivity is completely
determined by a single memory function $\chi.$ The method is used to study the
conductivity of the Fr\"{o}hlich polaron. It is found that the present
approach results in a conductivity which is three times higher than the one
predicted by Kadanoff and differs from that of FHIP by a factor $2\hbar
\beta \omega_{LO}.$ Consequently we recover the result of Los'~\cite{cit:LOS}.
In order to elucidate the difference, we have reanalyzed the Boltzmann
equation used by Kadanoff and the approach used by FHIP. Whereas the
relaxation time approximation used by Kadanoff explicitly violates particle
number conservation, the method developed by FHIP does not. The FHIP
approximation however relies on an unphyiscal initial state for Feynman's
polaron model. We find that a slightly modified version of both, which amends
these two problems, accounts for their discrepancy. \newpage%

%

\appendix

\section{Truncated equation of motion\label{sec:AppTruncated}}

Multiplying the Liouville equation~(\ref{eq:Linear:Liouville}) with
$e\mathbf{p}/m$ and integrating out the momentum yields%
\begin{multline}
\int \frac{e\mathbf{p}}{m}\left(  \frac{\partial}{\partial t}+e\mathbf{E}%
\left(  t\right)  \cdot \frac{d}{d\mathbf{p}}\right)  f\left(  \mathbf{p}%
,t\right)  \mathrm{d}\mathbf{p}=\sum_{\mathbf{k}}\frac{2\left \vert
\gamma \left(  k\right)  \right \vert ^{2}}{\hbar^{2}}\\
\times \int_{-\infty}^{t}\left(
\begin{array}
[c]{c}%
\int \frac{e\mathbf{p}}{m}f\left(  \mathbf{p-\hbar \mathbf{k}},s\right)  \left(
\begin{array}
[c]{c}%
\left(  n_{B}\left(  \omega_{k}\right)  +1\right)  \cos \left(  \left(
t-s\right)  \left(  \mathbf{k\cdot}\frac{\mathbf{p-}\frac{\hbar \mathbf{k}}{2}%
}{m}+\omega_{k}\right)  \right) \\
+n_{B}\left(  \omega_{k}\right)  \cos \left(  \left(  t-s\right)  \left(
\mathbf{k\cdot}\frac{\mathbf{p-}\frac{\hbar \mathbf{k}}{2}}{m}-\omega
_{k}\right)  \right)
\end{array}
\right)  \mathrm{d}\mathbf{p}\\
-\int \frac{e\mathbf{p}}{m}f\left(  \mathbf{p},s\right)  \left(
\begin{array}
[c]{c}%
\left(  n_{B}\left(  \omega_{k}\right)  +1\right)  \cos \left(  \left(
t-s\right)  \left(  \mathbf{k\cdot}\frac{\mathbf{p+}\frac{\hbar \mathbf{k}}{2}%
}{m}+\omega_{k}\right)  \right) \\
+n_{B}\left(  \omega_{k}\right)  \cos \left(  \left(  t-s\right)  \left(
\mathbf{k\cdot}\frac{\mathbf{p+}\frac{\hbar \mathbf{k}}{2}}{m}-\omega
_{k}\right)  \right)
\end{array}
\right)  \mathrm{d}\mathbf{p}%
\end{array}
\right)  \mathrm{d}s.
\end{multline}
Taking the expression~(\ref{eq:Generic:defineJ}) for the current density into
account, the left hand side can directly be calculated. After the substitution
$\mathbf{p-\hbar \mathbf{k\rightarrow p}}$ in the first term on the right hand
side, one is left with%
\begin{equation}
\frac{d\mathbf{J}(t)}{dt}-\frac{e^{2}}{m}\mathbf{E(}t\mathbf{)}=\sum
_{\mathbf{k}}\frac{2\left \vert \gamma \left(  k\right)  \right \vert ^{2}}%
{\hbar}\mathbf{k}\frac{e}{m}\int_{-\infty}^{t}\int f\left(  \mathbf{p}%
,s\right)  \left(
\begin{array}
[c]{c}%
\left(  n_{B}\left(  \omega_{k}\right)  +1\right)  \cos \left(  \left(
t-s\right)  \left(  \frac{\mathbf{k\cdot p}}{m}+\frac{\hbar k^{2}}{2m}%
+\omega_{k}\right)  \right) \\
+n_{B}\left(  \omega_{k}\right)  \cos \left(  \left(  t-s\right)  \left(
\frac{\mathbf{k\cdot p}}{m}+\frac{\hbar k^{2}}{2m}-\omega_{k}\right)  \right)
\end{array}
\right)  \mathrm{d}\mathbf{p}\mathrm{d}s.
\end{equation}
Using the $\mathbf{k}\leftrightarrow-\mathbf{k}$ symmetry results in%
\begin{multline}
\frac{d\mathbf{J}(t)}{dt}-\frac{e^{2}}{m}\mathbf{E(}t\mathbf{)}=-\sum
_{\mathbf{k}}\frac{2\left \vert \gamma \left(  k\right)  \right \vert ^{2}}%
{\hbar}\mathbf{k}\int_{-\infty}^{t}\left(
\begin{array}
[c]{c}%
n_{B}\left(  \omega_{k}\right)  \sin \left(  \left(  \frac{\hbar k^{2}}%
{2m}-\omega_{k}\right)  \left(  t-s\right)  \right) \\
+\left(  n_{B}\left(  \omega_{k}\right)  +1\right)  \sin \left(  \left(
\frac{\hbar k^{2}}{2m}+\omega_{k}\right)  \left(  t-s\right)  \right)
\end{array}
\right) \label{eq:AppTrunc:dJ/dt_full}\\
\times \frac{e}{m}\int f\left(  \mathbf{p},s\right)  \sin \left(  \frac
{\mathbf{k\cdot p}}{m}\left(  t-s\right)  \right)  \mathrm{d}\mathbf{p}%
\mathrm{d}s.
\end{multline}
Since the current density~(\ref{eq:Generic:defineJ}) is of order $\mathbf{E,}$
the dominant contribution in the last line of this equation is provided by the
small momenta. It thus seems reasonable to expand the sine function:%
\begin{align*}
\frac{e}{m}\int f\left(  \mathbf{p},s\right)  \sin \left(  \frac{\mathbf{k\cdot
p}}{m}\left(  t-s\right)  \right)  \mathrm{d}\mathbf{p}  &  =\frac{e}{m}\int
f\left(  \mathbf{p},s\right)  \left(  \frac{\mathbf{k\cdot p}}{m}\left(
t-s\right)  -\left(  \frac{\mathbf{k\cdot p}}{m}\right)  ^{3}\frac{\left(
t-s\right)  ^{3}}{6}+\cdots \right)  \mathrm{d}\mathbf{p}\\
&  =\left(  t-s\right)  \frac{\mathbf{k\cdot J}\left(  s\right)  }{m}%
-\frac{\left(  t-s\right)  ^{3}}{6}\frac{e}{m}\int f\left(  \mathbf{p}%
,s\right)  \left(  \frac{\mathbf{k\cdot p}}{m}\right)  ^{3}\mathrm{d}%
\mathbf{p}+\cdots.
\end{align*}

For general coupling strength $\gamma \left(  k\right)  $ and temperature, this
expansion seems not very useful. Indeed, the Wigner function broadens with
increasing temperature. Furthermore, for strong coupling the initial phonon
states are better described by a displaced and broadened Gaussian wave
functions, as shown in the derivation of the optical absorption of polarons
in~\cite{cit:DeFilippisetal2006,cit:KliminDevreese}. The change in the initial
phonon state will effect the influence
phase~\cite{cit:SBMinfluencefunctional2010} and consequently the self
energy~\cite{cit:SBreducedLiouville2013}. A dressed propagator will replace
the free particle propagator (\ref{eq:Generic:K0}). The extension of the
present result to strong coupling will be a topic of forthcoming work.

However, in the present paper we were mainly concerned with the discrepancy
between the FHIP result and the Kadanoff result for small electron-phonon
coupling and low temperature. In that case, neither the electron-phonon
coupling nor the temperature are able to broaden the distribution function
substantially. Therefore, for $\gamma \left(  k\right)  $ and $T$ sufficiently
small, one might truncate the expansion to the first moment, which results in
\begin{equation}
\frac{d\mathbf{J}(t)}{dt}-\frac{e^{2}}{m}\mathbf{E(}t\mathbf{)}\approx
-\sum_{\mathbf{k}}\frac{2\left \vert \gamma \left(  k\right)  \right \vert ^{2}%
}{\hbar}\mathbf{k}\int_{-\infty}^{t}\left(
\begin{array}
[c]{c}%
\left(  n_{B}\left(  \omega_{k}\right)  +1\right)  \sin \left(  \left(
t-s\right)  \left(  \frac{\hbar k^{2}}{2m}+\omega_{k}\right)  \right) \\
+n_{B}\left(  \omega_{k}\right)  \sin \left(  \left(  t-s\right)  \left(
\frac{\hbar k^{2}}{2m}-\omega_{k}\right)  \right)
\end{array}
\right)  \left(  t-s\right)  \frac{\mathbf{k\cdot J}\left(  s\right)  }%
{m}\mathrm{d}s. \label{eq:AppTruncated:dJ/dt}%
\end{equation}
Note that one can systematically improve the result~\cite{cit:Polkovnikov2010}
by the equations of motion for the higher moments.

\section{Relaxation time approximation\label{sec:AppendixRelaxation}}

The purpose of this Appendix is to explain the discrepancy
in~(\ref{eq:Frohlich:sigma_onze_Kad_FHIP}) by a factor of 3 between the DC
conductivity of the Fr\"{o}hlich polaron which we derived
in~(\ref{eq:Frohlich:sigmaDC}), as compared to the Kadanoff
result~\cite{cit:Kadanoff1963}. We thus consider the linearized Liouville
equation~(\ref{eq:Linear:Liouville}) for the reduced Wigner function. Using
$\int_{-\infty}^{t}\cos \left(  \left(  t-s\right)  a\right)  \mathrm{d}%
s=\pi \delta \left(  a\right)  $ one easily derives that its stationary version
is a Boltzmann equation%
\begin{equation}
e\mathbf{E}\cdot \frac{df\left(  \mathbf{p}\right)  }{d\mathbf{p}}=-\Pi \left(
\mathbf{p}\right)  f\left(  \mathbf{p}\right)  +\int \Pi \left(  \mathbf{p}%
+\hbar \mathbf{k}\rightarrow \mathbf{p}\right)  f\left(  \mathbf{p+\hbar
\mathbf{k}}\right)  d\mathbf{k,} \label{eq:AppendixRelaxation:Liouville}%
\end{equation}
with $\Pi \left(  \mathbf{p}+\hbar \mathbf{k}\rightarrow \mathbf{p}\right)  $ and
$\Pi \left(  \mathbf{p}\right)  $ defined in~(\ref{eq:Generic:PI(...)_define})
and~(\ref{eq:Generic:PI_total_define}).

Because the unperturbed reduced Wigner distribution function at sufficiently
low temperature peaks around $\mathbf{p}=0,$ one might argue that the dominant
term in the right hand side is given by $-f\left(  \mathbf{p}\right)
\lim_{\mathbf{p\rightarrow0}}\Pi \left(  \mathbf{p}\right)  ,$ which gives rise
to a relaxation time approximation (RTA):
\[
e\mathbf{E}\cdot \frac{df\left(  \mathbf{p}\right)  }{d\mathbf{p}}\approx
-\frac{f\left(  \mathbf{p}\right)  }{\tau}\text{ with }\tau=\frac{1}%
{\lim_{\mathbf{p\rightarrow0}}\Pi \left(  \mathbf{p}\right)  }.
\]
The first moment of this equation with respect to $\mathbf{p,}$
taking~(\ref{eq:Generic:defineJ}) into account, then immediately leads to%
\[
\mathbf{J}=\lim_{\mathbf{p}\rightarrow0}\frac{e^{2}/m}{\Pi \left(
\mathbf{p}\right)  }\mathbf{E}\text{ hence }\underset{\text{{\small RTA}}%
}{\sigma_{DC}}=\lim_{\mathbf{p}\rightarrow0}\frac{e^{2}/m}{\Pi \left(
\mathbf{p}\right)  }.
\]

For the Fr\"{o}hlich polaron, with the constant frequency $\omega_{k}%
=\omega_{LO}$ and the electron-phonon coupling~(\ref{eq:Frohlich:gamma(k)}),
the corresponding function $\Pi_{\text{Fr\"{o}hlich}}\left(  \mathbf{p}%
\right)  $ can easily be calculated in closed form:%
\[
\Pi_{\text{Fr\"{o}hlich}}\left(  \mathbf{p}\right)  =2\alpha \omega_{LO}%
\frac{\sqrt{2m\hbar \omega_{LO}}}{p}\left(
\begin{array}
[c]{c}%
\left(  n_{B}(\omega_{LO})+1\right)  \Theta \left(  \hbar \omega_{LO}%
<\frac{p^{2}}{2m}\right)  \operatorname{arccosh}\frac{p}{\sqrt{2m\hbar
\omega_{LO}}}\\
+n_{B}(\omega_{LO})\operatorname{arcsinh}\frac{p}{\sqrt{2m\hbar \omega_{LO}}}%
\end{array}
\right)  ,
\]
This simple relaxation time approximation thus immediately gives the Kadanoff
conductivity for the Fr\"{o}hlich polaron:
\[
\underset{\text{Kadanoff}}{\sigma_{DC}}=\lim_{\mathbf{p}\rightarrow0}%
\frac{e^{2}/m}{\Pi_{\text{Fr\"{o}hlich}}\left(  \mathbf{p}\right)  }%
\approx \frac{1}{2}\frac{e^{2}}{m\alpha \omega_{LO}}e^{\beta \hbar \omega_{LO}}.
\]

However, the neglect of the integral term
in~(\ref{eq:AppendixRelaxation:Liouville}) is an unwarranted approximation,
essentially because it violates the particle number conservation. Indeed,
consider the first moment of (\ref{eq:AppendixRelaxation:Liouville}) with
respect to $\mathbf{p:}$
\[
e\mathbf{E}=\int \mathbf{p}\Pi \left(  \mathbf{p}\right)  f\left(
\mathbf{p}\right)  d\mathbf{p}-\int \int \mathbf{p}\Pi \left(  \mathbf{p}%
+\hbar \mathbf{k}\rightarrow \mathbf{p}\right)  f\left(  \mathbf{p+\hbar
\mathbf{k}}\right)  d\mathbf{kdp.}%
\]
By the substitution $\mathbf{p+\hbar \mathbf{k\rightarrow p}}$ in the last
term, interchanging $\mathbf{k\leftrightarrow-k}$ and using the
definition~(\ref{eq:Linear:chi}), the terms in $\Pi \left(  \mathbf{p}\right)
$ cancel against each other, and one is left with
\[
eE=-\mathbf{1}_{E}\cdot \int \int \mathbf{\hbar \mathbf{k}}\Pi \left(
\mathbf{p}\rightarrow \mathbf{p+\hbar \mathbf{k}}\right)  f\left(
\mathbf{p}\right)  d\mathbf{kdp,}%
\]
which shows that the in-scattering rate can not be neglected.

At sufficiently low temperature, the distribution function peaks at
$\mathbf{\bar{p}=}m\mathbf{J}/e$ which is indeed near $\mathbf{p=0}$ since
$\mathbf{\bar{p}}\propto \mathbf{E}\rightarrow \mathbf{0.}$ Replacing $f\left(
\mathbf{p}\right)  $ by $\delta \left(  \mathbf{p-}m\mathbf{J}/e\right)  $ then
gives%
\begin{equation}
eE=-\mathbf{1}_{E}\cdot \int \mathbf{\hbar \mathbf{k}}\Pi \left(  \frac
{m\mathbf{J}}{e}\rightarrow \frac{m\mathbf{J}}{e}\mathbf{+\hbar \mathbf{k}%
}\right)  d\mathbf{k.}%
\end{equation}

For the Fr\"{o}hlich polaron~(\ref{eq:Frohlich:gamma(k)}), the evaluation of
this integral is elementary and results in:%
\begin{multline}
eE\mathbf{=}m\omega_{LO}\alpha \sqrt{2}\sqrt{\frac{\hbar \omega_{LO}}{m}}%
\frac{2e^{2}}{mJ^{2}}\\
\times \left(
\begin{array}
[c]{c}%
\left(  n_{B}\left(  \omega_{LO}\right)  +1\right)  \Theta \left(  \hbar
\omega_{LO}<\frac{mJ^{2}}{2e^{2}}\right)  \left(  \frac{\sqrt{m}J}{\sqrt{2}%
e}\sqrt{\frac{mJ^{2}}{2e^{2}}-\hbar \omega_{LO}}+\hbar \omega_{LO}%
\operatorname{arccosh}\left(  \frac{J}{e}\frac{\sqrt{m}}{\sqrt{2\hbar
\omega_{LO}}}\right)  \right) \\
+n_{B}\left(  \omega_{LO}\right)  \left(  \frac{\sqrt{m}J}{\sqrt{2}e}%
\sqrt{\frac{mJ^{2}}{2e^{2}}+\hbar \omega_{LO}}-\hbar \omega_{LO}%
\operatorname{arcsinh}\left(  \frac{J}{e}\frac{\sqrt{m}}{\sqrt{2\hbar
\omega_{LO}}}\right)  \right)  \allowbreak
\end{array}
\right)  .
\end{multline}
Keeping linear response in mind, it is obvious that this expression is only
needed to first order in $J=O\left(  E\right)  ,$ such that the emission term
does not contribute at sufficiently low temperature. The result is%
\begin{equation}
eE\mathbf{=}\frac{2}{3}m\omega_{LO}\alpha n_{B}\left(  \omega_{LO}\right)
\frac{J}{e}+O\left(  J^{3}\right)  ,
\end{equation}
which is fully consistent with the conductivity~(\ref{eq:Frohlich:sigmaDC})
derived above.

\section{FHIP with distribution function\label{sec:AppendixFHIP}}

In this section we present a calculation in the spirit of the FHIP
approximation but using our phase space approach. It was shown
in~\cite{cit:SBreducedLiouville2013} how the path integral for the reduced
Wigner function leads to the Liouville equation~(\ref{eq:Generic:Liouville}).
The path integral for the reduced Wigner function is just the Weyl transform
of the path integral for the density matrix used by FHIP. The basic approach
in FHIP is to expand the action around Feynman's linear polaron model, rather
than around the free particle. In terms of the distribution function this
means that
\begin{equation}
f(\mathbf{p,}t)=f_{0}(\mathbf{p,}t)+f_{1}(\mathbf{p,}t),
\end{equation}
where $f_{0}$ is a variational time dependent Wigner function which can be
found by propagating the initial distribution along a certain, so far free to
choose, linear model. Similar as for the linear response at weak coupling
(i.e., to first order in the deviation from the free particle), we now
consider linear response to first order in the deviation from the Feynman
polaron model, which means that%
\begin{equation}
\left(  \frac{\partial}{\partial t}+e\mathbf{E\cdot}\nabla_{p}\right)
f_{1}(\mathbf{p,}t)=g_{0}(\mathbf{p},t),
\end{equation}
where $g_{0}(\mathbf{p},t),$ apart from the time evolution of $f_{0},$ is the
right hand side of~(\ref{eq:Linear:Liouville}) with $f$ replaced by $f_{0}:$%
\begin{multline}
g_{0}(\mathbf{p},t)=-\left(  \frac{\partial}{\partial t}+e\mathbf{E(}%
t\mathbf{)}\cdot \nabla \right)  f_{0}\left(  \mathbf{p},t\right)  \\
+\sum_{\mathbf{k}}\frac{2\left \vert \gamma \left(  k\right)  \right \vert ^{2}%
}{\hbar^{2}}\int_{-\infty}^{t}\left(
\begin{array}
[c]{c}%
f_{0}\left(  \mathbf{p+\hbar \mathbf{k}},s\right)  \left(
\begin{array}
[c]{c}%
\left(  n_{B}\left(  \omega_{k}\right)  +1\right)  \cos \left(  \left(
t-s\right)  \left(  \mathbf{k\cdot}\frac{\mathbf{p+}\frac{\hbar \mathbf{k}}{2}%
}{m}-\omega_{k}\right)  \right)  \\
+n_{B}\left(  \omega_{k}\right)  \cos \left(  \left(  t-s\right)  \left(
\mathbf{k\cdot}\frac{\mathbf{p+}\frac{\hbar \mathbf{k}}{2}}{m}+\omega
_{k}\right)  \right)
\end{array}
\right)  \\
-f_{0}\left(  \mathbf{p},s\right)  \left(
\begin{array}
[c]{c}%
\left(  n_{B}\left(  \omega_{k}\right)  +1\right)  \cos \left(  \left(
t-s\right)  \left(  \mathbf{k\cdot}\frac{\mathbf{p+}\frac{\hbar \mathbf{k}}{2}%
}{m}+\omega_{k}\right)  \right)  \\
+n_{B}\left(  \omega_{k}\right)  \cos \left(  \left(  t-s\right)  \left(
\mathbf{k\cdot}\frac{\mathbf{p+}\frac{\hbar \mathbf{k}}{2}}{m}-\omega
_{k}\right)  \right)
\end{array}
\right)
\end{array}
\right)  \mathrm{d}s.\label{eq:AppendixFHIP_g0}%
\end{multline}
The time dependence of the distribution function $f_{1}$ follows the classical
equation of motion, and consequently%
\[
f_{1}(\mathbf{p,}t)=\int_{-\infty}^{t}g_{0}\left(  \mathbf{p}-\int_{t^{\prime
}}^{t}e\mathbf{E}(s)\mathrm{d}s,t^{\prime}\right)  \mathrm{d}t^{\prime}.
\]
Because of the particle number conservation of the trial distribution, and due
to the linearity of the classical equation of motion, the expected current
density of the perturbation around the model becomes%
\begin{equation}
\mathbf{J}_{1}(t)=\frac{e}{m}\int \mathbf{p}f_{1}(\mathbf{p,}t)\mathrm{d}%
\mathbf{p}=\frac{e}{m}\int_{-\infty}^{t}\int \mathbf{p}g_{0}(\mathbf{p}%
,t^{\prime})\mathrm{d}\mathbf{p}\mathrm{d}t^{\prime}%
.\label{eq:AppendixFHIP_currentcorrection}%
\end{equation}
The total current density is consequently given by%
\[
\mathbf{J}(t)=\mathbf{J}_{0}(t)+\mathbf{J}_{1}(t),
\]
where $\mathbf{J}_{0}(t)$ is the current density of the model distribution
function. In terms of Feynman's variational parameters $w$ and $v,$ Feynman's
model distribution function reads%
\[
f_{0}(\mathbf{p,}t)=\left(  \frac{\beta}{2m\pi}\right)  ^{3/2}\exp \left(
-\frac{\beta}{2m}\left(  \mathbf{p}-\frac{w^{2}}{v^{2}}\int_{-\infty}%
^{t}e\mathbf{E(}s\mathbf{)}\mathrm{d}s-\frac{v^{2}-w^{2}}{v^{2}}\int_{-\infty
}^{t}e\mathbf{E(}s\mathbf{)}\cos v(t-s)\mathrm{d}s\right)  ^{2}\right)  ,
\]
provided we assume the model to be initially in canonical equilibrium at an
effective temperature equal to the real temperature $\beta^{-1}.$ At this
point the present discussion differs from that of FHIP, where the initial
state of the model is assumed to be a product state of the oscillators with
the particle. It is argued by FHIP that the product state ansatz is admissible
because \textit{\textquotedblright... }\emph{In the past only the oscillators
were in thermal equilibrium at }$\beta^{-1}.$\emph{ As a result of the
coupling the system will come very quickly to thermal equilibrium at the same
temperature.~}\cite{cit:FHIP1962}\textit{\textquotedblright \ }Although this
might be true for the real system, it does not apply to the model. Because of
the linearity of the model it will never thermalize. Consequently, the reduced
model distribution function will endlessly oscillate even in the absence of an
electric field. In contrast, the present model distribution is the exact
stationary distribution of the reduced Liouville equation in the absence of an
electric field~\cite{cit:SBreducedLiouville2013}. It should however also be
noted that, as a consequence of the same linearity, the expected model current
density
\[
\mathbf{J}_{0}(t)=\frac{w^{2}}{v^{2}}\int_{-\infty}^{t}\frac{e^{2}%
\mathbf{E(}s\mathbf{)}}{m}\mathrm{d}s+\frac{v^{2}-w^{2}}{v^{2}}\int_{-\infty
}^{t}\frac{e^{2}\mathbf{E(}s\mathbf{)}}{m}\cos v(t-s)\mathrm{d}s,
\]
is not affected by the change in initial state, in contrast to the correction
$\mathbf{J}_{1}(t)$. From the definition~(\ref{eq:Generic:definesigma}) of the
conductivity, we furthermore find the following expression for the Laplace
transform $%
\mathcal{L}%
\left(  \sigma_{0},\Omega \right)  $ of the model conductivity
\[%
\mathcal{L}%
\left(  \sigma_{0},\Omega \right)  =\frac{e^{2}}{m}\left(  \frac{w^{2}}{v^{2}%
}\frac{1}{\Omega}+\frac{v^{2}-w^{2}}{v^{2}}\frac{\Omega}{v^{2}+\Omega^{2}%
}\right)  .
\]

The first order correction $\mathbf{J}_{1}(t)$ consists of two parts, one that
scales with the coupling constant and one that does not. The latter one is
given by%
\begin{align*}
\mathbf{J}_{1,0}(t)  &  =-\int_{-\infty}^{t}\int e\mathbf{p}\left(
\frac{\partial}{\partial t}+e\mathbf{E(}t\mathbf{)\cdot}\nabla_{p}\right)
f_{0}(\mathbf{p,}t^{\prime})\mathrm{d}\mathbf{p}\mathrm{d}t^{\prime}.\\
&  =\frac{v^{2}-w^{2}}{v^{2}}\left[  \int_{-\infty}^{t}\frac{e^{2}%
\mathbf{E(}s\mathbf{)}}{m}\mathrm{d}s-\int_{-\infty}^{t}\frac{e^{2}%
\mathbf{E(}s\mathbf{)}}{m}\cos v(t-s)\mathrm{d}s\right]  .
\end{align*}
The coupling dependent part leads to%
\[
\mathbf{J}_{1,1}(t)=\frac{e}{m}\int_{-\infty}^{t}\mathrm{d}t^{\prime}%
\int_{-\infty}^{t^{\prime}}\mathrm{d}s\sum_{\mathbf{k}}\frac{2\left \vert
\gamma \left(  k\right)  \right \vert ^{2}}{\hbar}\mathbf{k}\int f_{0}\left(
\mathbf{p},s\right)  \left(
\begin{array}
[c]{c}%
\left(  n_{B}\left(  \omega_{k}\right)  +1\right)  \cos \left(  \left(
t^{\prime}-s\right)  \left(  \frac{\mathbf{k\cdot p}}{m}+\frac{\hbar k^{2}%
}{2m}+\omega_{k}\right)  \right) \\
+n_{B}\left(  \omega_{k}\right)  \cos \left(  \left(  t^{\prime}-s\right)
\left(  \frac{\mathbf{k\cdot p}}{m}+\frac{\hbar k^{2}}{2m}-\omega_{k}\right)
\right)
\end{array}
\right)  \mathrm{d}\mathbf{p,}%
\]
which within linear response, hence up to $O\left(  E\right)  ,$ simplifies to%
\[
\mathbf{J}_{1,1}(t)=-\int_{-\infty}^{t}\mathrm{d}t^{\prime}\int_{-\infty
}^{t^{\prime}}\mathrm{d}s\chi_{\beta}\left(  t^{\prime}-s\right)
\mathbf{J}_{0}(s),
\]
with
\[
\chi_{\mathbf{\beta}}(t)=t\sum_{\mathbf{k}}\frac{2\left \vert \gamma \left(
k\right)  \right \vert ^{2}}{3\hbar}\frac{\mathbf{k}^{2}}{m}\left(
\begin{array}
[c]{c}%
\left(  n_{B}\left(  \omega_{k}\right)  +1\right)  \sin \left(  t\left(
\frac{\hbar \mathbf{k}^{2}}{2m}+\omega_{k}\right)  \right) \\
+n_{B}\left(  \omega_{k}\right)  \sin \left(  t\left(  \frac{\hbar
\mathbf{k}^{2}}{2m}-\omega_{k}\right)  \right)
\end{array}
\right)  \exp \left(  -\frac{\mathbf{k}^{2}}{2m\beta}t^{2}\right)  .
\]
Note that $\lim_{\beta \rightarrow \infty}\chi_{\mathbf{\beta}}(t)=\chi(t),$
where $\chi(t)$ is the memory function obtained by truncating the equation of
motion for the current density, as explained in
appendix~\ref{sec:AppTruncated}. Consequently the low temperature, linear
response, current density up to first order around the Feynman polaron model
is%
\[
\mathbf{J}(t)=\int_{-\infty}^{t}\frac{e^{2}\mathbf{E(}s\mathbf{)}}%
{m}\mathrm{d}s-\int_{-\infty}^{t}\mathrm{d}t^{\prime}\int_{-\infty}%
^{t^{\prime}}\mathrm{d}s\chi \left(  t^{\prime}-s\right)  \mathbf{J}_{0}(s).
\]
Hence the Laplace transform $%
\mathcal{L}%
\left(  \sigma,\Omega \right)  $ of the conductivity reads%
\[%
\mathcal{L}%
\left(  \sigma,\Omega \right)  =%
\mathcal{L}%
\left(  \sigma_{0},\Omega \right)  +%
\mathcal{L}%
\left(  \sigma_{1},\Omega \right)  ,
\]
where the correction to the model conductivity $\sigma_{1}$ is given by
\[%
\mathcal{L}%
\left(  \sigma_{1},\Omega \right)  =\frac{e^{2}}{m}\frac{v^{2}-w^{2}}%
{\Omega \left(  v^{2}+\Omega^{2}\right)  }-\frac{%
\mathcal{L}%
\left(  \sigma_{0},\Omega \right)
\mathcal{L}%
\left(  \chi,\Omega \right)  }{\Omega}.
\]

A more accurate conductivity can be found using the standard resummation
argument
\[%
\mathcal{L}%
\left(  \sigma,\Omega \right)  =%
\mathcal{L}%
\left(  \sigma_{0},\Omega \right)  \left(  1+\frac{%
\mathcal{L}%
\left(  \sigma_{1},\Omega \right)  }{%
\mathcal{L}%
\left(  \sigma_{0},\Omega \right)  }\right)  \approx \frac{%
\mathcal{L}%
\left(  \sigma_{0},\Omega \right)  }{1-\frac{%
\mathcal{L}%
\left(  \sigma_{1},\Omega \right)  }{%
\mathcal{L}%
\left(  \sigma_{0},\Omega \right)  }},
\]
that is expression (38) in FHIP. \ Consequently the DC-conductivity for the
optical Fr\"{o}hlich polaron reads
\[
\sigma_{DC}=\lim_{\Omega \rightarrow0}%
\mathcal{L}%
\left(  \sigma,\Omega \right)  =\frac{w^{2}}{v^{2}}\frac{3e^{2}}{2\alpha
m\omega_{LO}n_{B}(\omega_{LO})}.
\]
Moreover, since $v^{2}/w^{2}=m^{\ast}/m$ \cite{cit:FHIP1962} we have%
\[
\sigma_{DC}=\frac{3e^{2}}{2\alpha m^{\ast}\omega_{LO}n_{B}(\omega_{LO}%
)}\approx \frac{3e^{2}}{2\alpha m^{\ast}\omega_{LO}}e^{\beta \hbar \omega_{LO}},
\]
consistent with our result~(\ref{eq:Frohlich:sigmaDC}).

\begin{acknowledgments}
The authors thank Prof. J.T. Devreese for many stimulating discussions.
\end{acknowledgments}

\end{document}